\documentclass[aps]{revtex4}

\usepackage{graphicx}
\usepackage{amsmath}
\usepackage{caption}
\def\be{\begin{equation}}
\def\ee{\end{equation}}
\def\bi{\bibitem}

\begin{document}

\title{The Pareto Principle in Sports and Economics based on Runs Scored by Batters in the Indian Premier League}

\author{Soumendra Nath Ruz}
\email{ruzfromju@gmail.com}
\affiliation{Department of Physics, Ramananda Centenary College, Laulara, Purulia 723151, India}
\author{Asim Ghosh }
\email{asimghosh066@gmail.com}
\affiliation{Department of Physics, Raghunathpur College, Raghunathpur 723133, India}

\begin{abstract}
  The analysis of income and wealth inequality is often constrained by the lack of reliable data. In this work, we introduce a proxy-based approach in which sports performance data are used to mimic economic distributions. In particular, the total runs scored by a batter in a single T20 season are treated as a proxy for annual income, while the cumulative runs scored across all seasons are taken to represent individual wealth. Using run distributions from the Indian Premier League (IPL), we explore how inequality evolves and estimate its upper limits in the absence of policy intervention. Our findings show that the inequality caused by seasonal runs is very similar to the highest levels of income inequality we see in real life. In addition, the distribution of cumulative runs evolves over time and gradually approaches a limiting value consistent with global patterns of wealth inequality, in line with the Pareto principle. In general, this study shows that proxy systems can capture essential features of economic inequality and offer a useful way to understand its inherent limits.
\end{abstract}

\maketitle

\section{Introduction}
Income and wealth inequality represent significant socio-economic challenges worldwide. Inequality is closely linked to social and behavioral structures, particularly because humans are naturally motivated to achieve high social status. Despite various efforts to combat this, inequality remains a persistent issue. Pervasive inequality profoundly impacts social structure and stability. Research indicates that high inequality leads to increased crime, lower social trust, and, in many cases, worsening environmental sustainability  \cite{1a, 1b, 1c, 1d, 1e, 1f}. Such disparities inhibit social progress and significantly undermine the overall well-being of the population. Furthermore, recent literature surveys suggest that increasing income inequality diminishes social mobility and curtails equality of opportunity, subsequently inhibiting long-term growth by preventing the accumulation of human capital \cite{1g, 1h, 1i, 1j, 15c}. Therefore, governments implement strategic public policies like progressive taxation, robust labor rights protection, and progressive public spending to reduce inequality to a manageable level. In addition, governments utilize targeted welfare schemes, financial assistance programs, and significant investments in human capital—such as education and healthcare—to ensure an equitable distribution of resources and opportunities \cite{1a1,1a2}.

Thus, the study of income and wealth inequality i.e, an uneven distribution of earnings, assets, and opportunities among individuals or households, becomes a critical issue for a nation's well-being. Generally, income includes earnings from wages, salaries, interest from savings accounts, dividends from shares of stock, rent, and profits from selling something for more than you paid for it \cite{1g}. It is the earnings from the above-mentioned sources for a specific duration of time. This period may be in hours, days, months, or years. However, with income, we also have expenses to maintain our social life. On the other hand, wealth is the total value of assets owned minus liabilities at a specific point in time, representing accumulated net worth. While income funds current lifestyle and enables wealth accumulation, wealth itself provides security and generates future income. If income exceeds expenses, wealth increases over time; otherwise, it decreases.

In general, the inequality is measured either using graphical descriptions or various statistical indices. The Lorenz Curve is a primary graphical tool, while the Gini index, Palma Ratio (top 10\% vs. bottom 40\%), Theil index, and Atkinson index are numerical indicators \cite{1, 2, 2a, 2b}. Among these measures, the Gini index is the most widely used, ranging from 0 (perfect equality) to 1 (perfect inequality). In practice, values of 0 or 1 are rarely observed; instead, the Gini index typically lies between these extremes. Another index, the Kolkata index or $k$-index (also derived from the Lorenz curve), has been used recently \cite{Ghosh14}. The value of the $k$-index can vary from 0.5 to 1. A value of 0.5 indicates perfect equality, whereas 1 indicates absolute inequality. When the index value is $k$, it implies that the $k$ fraction of people possesses the $1-k$ fraction of wealth when it is applied to wealth inequality. Studies show that the upper limit of inequality depends on the extent of policy intervention, and in the absence of such measures, it generally reaches an upper limiting value \cite{2c1, 2c2, 2c3, 2c4, 2c5, 2c6}.  

Accurate measurement of household income and wealth data that are highly private is very difficult due to under-reporting, low response rates during household surveys, and participation of a large section of people in various informal work \cite{2d1, 2d2}. So, economists combine multiple data sources to estimate income inequality \cite{3a, 3a2, 3, 3b, 3b1}, while wealth inequality remains harder to measure due to non-response during surveys, difficulty in asset valuation issues, and missing data on hidden assets \cite{14a}. Despite these challenges, it is crucial for understanding economic structure, leading to the use of integrated methods \cite{15a, 15b, 4, 4a, 4b, 4c, 4d, 4e, a5}. Although global income inequality data are regularly updated by multiple trusted sources (e.g., World Bank, OECD), wealth inequality data is published less frequently \cite{7, 7a, 7b, 7c,7d,7f, 8a, 8c}. Globally, income inequality (Gini index) ranges from about 0.24 to 0.63, with countries like South Africa, Namibia, and Brazil showing high inequality, while Slovakia, Slovenia, and Belarus exhibit low levels; India has relatively low income inequality (~0.26) \cite{7b, 8a}. In contrast, wealth inequality is generally higher (Gini ~0.5–0.8), reaching extreme levels in countries such as South Africa (~0.82), where the top 10\% hold about 85–86\% of wealth \cite {8a,8b, 9a}, and in some cases, including India and Norway, low income inequality coexists with high wealth disparity.

In the absence of accurate data, measured inequality may not accurately reflect the true ground reality \cite{5, 5a, 5b, 5c, 6, 6a}. Consequently, policymakers often struggle to design effective welfare measures to reduce inequality and balance wealth distribution. To address this, we use accessible proxy parameters to mimic actual income and wealth data, allowing us to analyze how inequality evolves over time. Our goal is to study this evolution and determine the realistic upper limit of inequality in the absence of mitigation measures, drawing on literature regarding similar socio-economic issues \cite{2c1, 2c2, 2c3, 2c4, 2c5}. Inequality indices such as the Gini index have been measured in various fields, including the health sector, traffic congestion, and sports, in addition to income and wealth \cite{m29,m30,m31,m32,m33}. In particular, these indices have been used to measure competitive balance in sports, which is fundamentally concerned with the outcomes of sporting competitions \cite{Kevin20}. In the article, ``Championship Inequality in Major League Sports Looks a Lot Like Income Inequality in the U.S. Economy,'' of  Washington Post, Christopher Ingraham highlighted similarities between sports outcomes and wealth distribution in the United States \cite{Washingtonpost}. 

In the present article, we have considered the runs scored by batters in T20 matches of the Indian Premier League (IPL) since 2008 as a measure to calculate income and wealth inequality. Specifically, runs made by a batter in an innings have been treated as income, and the cumulative runs he has made in various innings of the IPL as his wealth. All data related to the runs scored by a batter can be collected from the official IPL website \cite{10} and various other standard sports websites \cite{11, 11a}. The next section introduces the mathematical framework for calculating inequality. Section III provides a brief description of the IPL for readers unfamiliar with the league. Section IV justifies the comparison between income/wealth and the runs scored by batters, while detailing the methodology for calculating this inequality. Section V presents various plots analyzing the key features of these findings, and finally, Section VI discusses our study's findings.

%%%%%%%%%%%%%%%%%%%%%%%%%%%%%%%%%%%%%%%%%%%%%%%%%%%%%%%%%%%%%%%%%%%
\section{Measuring Inequality}

%%%%%%%%%%%%%%%%%%%%%%% First Figure %%%%%%%%%%%%%%%%%%%%%%%%
\begin{figure} [ht]
 \includegraphics[width = 0.7 \textwidth ]{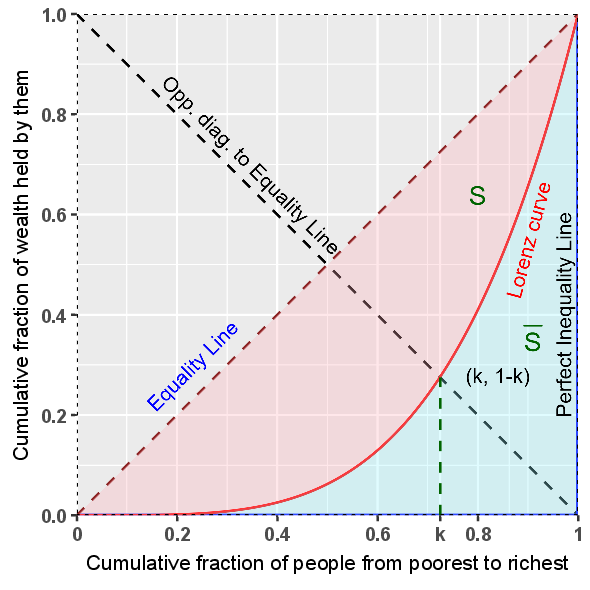}
 \caption {This plot depicts how to calculate the Gini index and the Kolkata index from a standard Lorenz plot. Here, the red solid line is the Lorenz curve for a certain distribution. The red dotted line represents the equality line, i.e. every fraction of population shares equal wealth, on the other hand the blue solid line is the perfect inequality line which represents only one person possesses all the wealth. $S$ is the pink shaded region between the equality line and the Lorenz curve. While $\bar{S}$ is the area between Lorenz curve and the perfect inequality line. In general, Gini index, $g = S/(S+\bar{S})$. Kolkata index $k$ is the value of the x-coordinate of the intersecting point between opposite diagonal of the equality line, i.e. $y=(1-x)$ and the Lorenz Curve. }
\label{Fig1}
\end{figure}
%%%%%%%%%%%%%%%%%%%%%%%%%%%%%%%%%%%%%%%%%%%%%%%%%%%%%%%%%%%%%%%%%%%%%%%%%%%%%%

Income and wealth inequality was systematically studied for the first time by Vilfredo Pareto \cite{5d1a}. The outcome of his study is popularly known as Pareto Principle which states that --approximately 80\% of outcomes come from 20\% of causes. Actually, Pareto found that nearly 80\% of Italy's land was owned by 20\% of the population. Later  Joseph M. Juran found that this law is roughly applicable to many other fields and on average 80\% of problems stem from 20\% of the causes \cite{11b}. Meanwhile, M.O. Lorenz expressed the inequality graphically, and this representation is known as the Lorenz curve \cite{aa8}. The Italian mathematician C Gini further developed an index  to indicate inequality and this is known as the Gini index \cite{11c}. It is calculated from the corresponding Lorenz curve. B k Chakraborty and his group found another inequality index named the Kolkata index, which directly depicts inequality in terms of the Pareto principle \cite{13a, 13b, 13c,13d,13e}. If $k$ is the Kolkata index, it informs that the $k$ fraction of people has a $(1-k)$ fraction of wealth. 

The estimation of the Gini index, $g$ and the Kolkata index, $k$ from the Lorenz curve has been shown in Fig.\ref{Fig1}, where the cumulative fraction of people according to wealth has been plotted along the $x$-axis and the wealth possessed by them along the $y$-axis. The red line indicates one such curve known as the Lorenz curve. Inequality is estimated by comparing its closeness with the equality line and the perfect inequality line. The equality line indicates that all wealth is distributed equally. The solid blue line is the perfect inequality line, which indicates that only one person possesses all the wealth of society.   

The Gini index ($g$) is the ratio of the area between the perfect equality line and the Lorenz curve, i.e., $(S)$(pink shaded area) and the total area under the perfect equality line, i.e., $(S+\bar{S})$. Mathematically,
\be 
g = \frac{S}{S+\bar{S}}. 
\label{eqn0}\ee 
In case of perfect equality, $S=0$ and $\bar{S} = \frac{1}{2}$, yielding $g=0$, while in case of perfect inequality, $S = \frac{1}{2} $ and $\bar{S} = 0$, yielding $g=1$. 

Kolkata Index ($k$) is the value of abscissa where the opposite diagonal of the equality line intersects the Lorenz curve. Observing Fig.\ref{Fig1}, one can notice that $k = 0.5$ at perfect equality and 1 at perfect inequality. The Gini index and the Kolkata index show a strong positive correlation between them \cite{13a, 13b, 13c,13d,13e}. When inequality is low, $k$ and $g$ approximately evolve according to the following relation with increasing inequality \cite{13d, 13f} 
\begin{equation}
 k = 0.5 + \frac{3}{8}g 
 \label{eqn2}
\end{equation}
The above linear relationship does not exist anymore when the inequality becomes high \cite{13d, 13e, 13f}.
Here we have briefly described the Lorenz curve and the measurement of the Gini and Kolkata index from it. We shall use these parameters to measure the inequality in the present article. Before going to the measurement of inequality, let us briefly discuss Indian Premier League (IPL) in the following section.

%%%%%%%%%%%%%%%%%%%%%%%%%%%%%%%%%%%%%%%%%%%%%%%%%
\section {A brief note on Indian Premier League (IPL)}
%%%%%%%%%%%%%%%%%%%%%%%%%%%%%%%%%%%%%%%%%%%%%%%%%
The Indian Premier League (IPL) is a well-known professional T20 cricket league in India, founded by the Board of Control for Cricket (BCCI) in 2007.  It is a type of club football played in Europe and Latin America. The first season was played in 2008, where 8 teams had participated. Each season is played every year. In 2011, two additional teams were included, and until now 10 teams are taking part in this event. Known for its high financial value and glamour, it features international superstars and emerging talent. Since its beginning in 2008, the IPL has become one of the most popular and financially successful cricket leagues in the world. Every franchise team includes both Indian and international players. Detailed rules and regulations, as well as information on past matches, can be found on the IPL's official website \cite{10} and other related websites \cite{11, 11a,12}. Here, we discuss the key features of this game. 

\begin{itemize}
    \item Teams are owned by business conglomerates and celebrities, operating on a business model that emphasizes commercial success and value of the team's brand. Annual player auctions allow teams to build balanced squads comprising top-tier international stars and domestic Indian players.
    \item Matches follow the official laws of cricket governed by the International Cricket Council. Each team plays 20 overs per innings with a 90-minute limit for each innings, including a 5-minute strategic timeout per innings. A match is played between two teams of 11 players each.
    \item The tournament typically follows a double round-robin format. For 10 participating teams, each team plays 14 matches in the league round.  Teams play matches at their designated home stadium and away at other franchises' stadiums. Each team earns 2 points for a win, 1 point for no result, and 0 for a loss. The top four teams in the points table qualify for the four-match playoff system (Qualifier 1, Eliminator, Qualifier 2, and Final) to determine the winner.
    \item The organizers distribute various awards at the end of each session to enhance the professionalism of the players. This practice motivates players to perform their best instead of doing average. In the process, it may improve the inequality because every player wants to be the best.
    
\end{itemize}
In other words, players from all over the world participate in this tournament with high enthusiasm. As it is of recent time and very popular, one can get all the information of the past matches from the websites depicted above. Most of the IPL seasons (2008–2025) were organized in India. However, due to pandemics or elections, venues have shifted to other countries from time to time. So in respect of players and venue it has an international aspect. For all these reasons, we have chosen the IPL tournament to study income and wealth inequality. 

In the next section, we have justified our claim about comparison of the runs scored by batters with the income and wealth inequality. At the same time, we have also discussed how inequality is measured for this particular case in terms of the Gini and Kolkata index.

%%%%%%%%%%%%%%%%%%%%%%%%%%%%%%%%%%%%%%%%%%%%%%%
\section{ Income and wealth inequality in terms of IPL data}
%%%%%%%%%%%%%%%%%%%%%%%%%%%%%%%%%%%%%%%%%%%%%
In the introduction, We have already discussed income and wealth. Income is necessary to meet our daily needs.  It acts as a foundation for achieving long-term goals, i.e., making wealth. On the other hand, wealth gives stability to a family/individual to sustain in the society. Wealthy people do not bother much, even if, at present, their income is low and has time to grow. 

let us try to find what income and wealth resemble in terms of runs scored by batters in the IPL. We discussed key features of IPL in the previous section. Around 10 teams are participating in recent IPL seasons. Each team competes in a round-robin in the league stages, followed by 4 playoffs and a final. Thus, every team plays several matches in a season. Every match consists of two innings. Batters from either teams get the chance to bat in one of the innings. As it is a 20 over innings match, batters of each team try to score more runs as quickly as possible. Moreover, various prizes lure the batters to play every innings in a spectacular way. In the process, a batter can play several innings if he stays on the team for the entire season. Each team generally also retains good batters in the next sessions. In the process, batters with good performance report play several innings.

We have segregated the runs scored by players in the IPL into two categories: 
\begin{itemize}
    \item Individual runs scored by players in each innings.
    \item Cumulative individual runs scored in all the past sessions till date.
\end{itemize}
Now, let us compare the role of these two types of runs in the career of a batter in the IPL. The cumulative run count gives a batter the stability to remain in the team for longer periods. If one batter has played earlier innings well, then his cumulative runs will be higher. It is well-known that a player cannot play well in all innings. At the same time, IPL matches are very competitive, and team management can replace any player in the team if his performance is not satisfactory. Suppose that a batter scores low in a particular innings, but his cumulative run count is good to date. Then, organizers will not remove him abruptly in the next match and will give him chances to do well in the future matches. However, if his cumulative runs are low and do not perform well in an innings, team management may remove him. Thus,  cumulative runs act as wealth for a player while the runs scored by him in each innings act like his income which grows wealth over time. If past income is good wealth gets accumulated and for a certain period even if income becomes low, he has time to survive. 

Now let us concentrate on how to calculate the inequality in individual runs scored by batters in each innings, i.e., income inequality. For this, one has to list which player has scored how many runs in each innings for a particular IPL season. At the same time, we have to keep in mind that the cumulative runs scored by the batters are related to wealth. So, while calculating income inequality, we have to judiciously discard cumulative runs. In this scenario, we have calculated the income inequality in the following way. 

\subsection{Income inequality in terms of runs scored in IPL}

Until now, 18 IPL sessions have been played. In every IPL session, so many matches have been played. All of these data can be retrieved from various trusted websites as mentioned earlier \cite{10, 11, 11a}. We have calculated income inequality session-wise starting from the first session played on 2008 to the 18-th session played on 2025. To calculate the inequality in income in a particular IPL session, we rearranged the runs scored by every batter per innings as follows. 

\subsubsection{Data Selection for income inequality}

All batters who have played at least one ball were listed innings-wise and match-wise for each session. Again, most of the batters have played multiple innings in each session. If we arrange the runs scored by each batter by their name for a particular session, then the batters who have played multiple innings will reflect cumulative run count which is related with wealth. To address this issue, we have labeled each batter with different numbers in each innings of every match. This numbering process continued serially from the first match to last of a particular session as per schedule. In the process, even if a batter has played multiple innings in different matches, they were tagged with different numbers associated with different innings. Thus, all the batters who have faced at least one ball in any innings, were tagged with a different number and the batters who have played more than one innings in a session have been treated as different entity while calculating income inequality. In this case each tagged number represents only a batter and not attached to their previous history, i.e. cumulative count. Thus we are in a position to calculate income inequality where no cumulative runs has been considered. In the next sub-subsection, we have described how we have plotted the Lorenz curve and calculated corresponding Gini and Kolkata index from it for each 18 IPL sessions played till now.

\subsubsection{Calculation of The Gini and Kolkata Index}

We have listed tagged numbers representing a batter and runs scored by them in each innings of every match for a particular season of IPL. Lorenz plot representing income inequality for each IPL session was plotted as follows:

\begin{itemize}
    \item Sort the list according to runs made by batters/innings in increasing order. Here batter with lowest runs will be first in the list and batters with highest runs will be last.
    \item Prepare a table of cumulative sum or running total from the above list. Cumulative sum is a sequence of partial sums of a given list of numbers, where each element in the new list is the sum of all preceding elements up to that position.
    \item Normalize the cumulative sum to 1. Here we shall calculate the Gini index in the scale from 0 to 1. If someone wants to calculate the Gini index in 0 to 100 scale, then the normalized sum should be 100. Here each entry in this normalized list is respective data points for the y-axis of the Lorenz plot. In the same way, if there are $N$ number of tagged numbers (representing batters) in the list we have prepared to calculate income inequality, then equally spaced by $\frac{1}{N}$ along the x-axis. Thus, the cumulative sum along the x-axis will also be 1.
    \item Plot the required Lorenz plot with above x and y-axis data. Now, as depicted in Section II, we have calculated the corresponding Gini and the Kolkata index from this Lorenz plot.
\end{itemize}

Following the above process, we have calculated the Gini and Kolkata index for each of the 18 IPL seasons played till now and listed in the Table \ref{Table1} given in the Appendix. The same has been plotted in Fig.\ref{fig2}. In this plot, the blue line represents the Gini index representing income inequality. It has been indicated by $g_{inc}$. The blue square points represent the value of the Gini index for the corresponding IPL sessions given along the x-axis. The brown line represents the same for the Kolkata index related to the income inequality. Next, we shall concentrate on the calculation of wealth inequality.

%%%%%%%%%%%%%%%%%% 2-nd Figure %%%%%%%%%%%%%%%%%%%
\begin{figure}[ht]
    \centering % Center the figure
    % Do not include the .eps extension in the file name
    \includegraphics[width = 1.0 \textwidth]{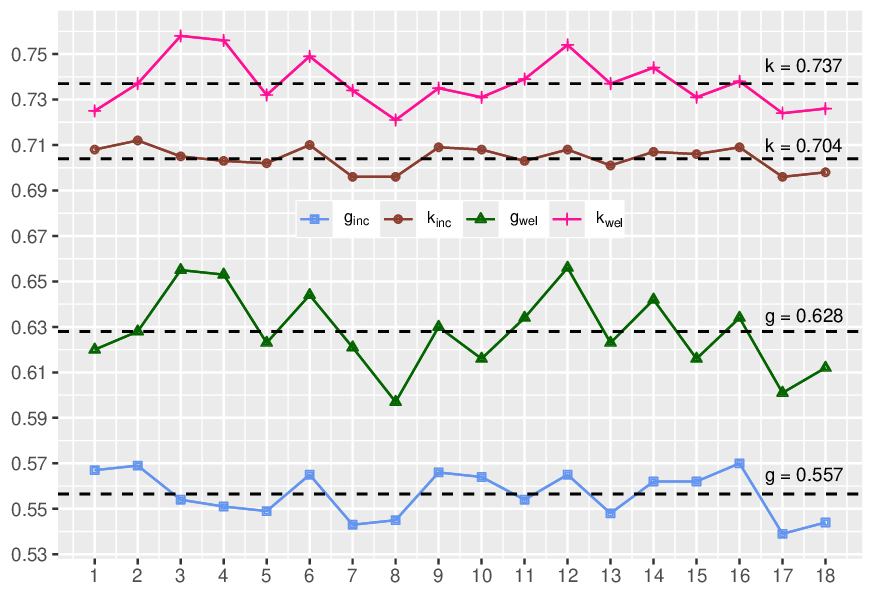} 
    \caption{ Variation of income inequality with each season and variation of the wealth inequality considering cumulative runs up to 1 season with each season in terms of the Gini and the Kolkata index. Here, $g_{inc}$ and
    $k_{inc}$ represents the Gini and Kolkata index for income inequality. While $g_{cum}$ and $k_{cum}$ represent the same for wealth
    inequality up to 1 season. The numbers along x-axis represent the respective IPL seasons. The numbers along y-axis represent the value of inequality. There are 18 points for each line plot. Each of these points indicates how inequality changes with respective sessions. It shows that wealth inequality is always higher than income inequality. The black dotted lines shows the average value of inequality. The average value of income inequality in terms of the Gini index is around 0.56 and in terms of the Kolkata index is 0.7. This means, in general 70\% runs are made by 30\% batters in each innings. Again average value of the wealth inequality for cumulative runs counted up to 1 session is 0.63 in Gini index and 0.74 in Kolkata index. This indicates, on an average, 73\% of the total runs scored in each IPL session are made by 27\% batters. }
    \label{fig2}
    
\end{figure}

%%%%%%%%%%%%%%%%% end of 2-nd figure %%%%%%%%%%

\subsection{Wealth Inequality in terms of runs scored in IPL}

We have already mentioned that any cumulative runs scored by the batters is related with his wealth. So, one can calculate wealth inequality considering the cumulative runs scored in any 1 session, or any 2 session and so on. In the maximum case, we can consider cumulative runs in all 18 session played till now for this particular case. Now, the wealth inequality for cumulative runs scored per 1 session can be obtained as follows:

\begin{itemize}
    \item  Every batter has different name which identifies them. So, in case of wealth inequality, there is no need to tag the batters by numbers as was done in income inequality. List the batters with the cumulative runs they have scored in an entire session.
    \item Sort the list in accordance to cumulative runs made in increasing order and prepare a table of cumulative sum from it as in the previous case. Normalize the cumulative sum table to 1. At this point each element in the table is the required y-coordinates of the Lorenz plot. If the are $N$ players in the cumulative list, then, as in the previous case $\frac{1}{N}$ equally spaced points will represent the required points along the x-axis of the Lorenz plot.
    \item Calculate corresponding Gini index and the Kolkata index from the above Lorenz plot as in case of wealth inequality.
    
\end{itemize}

We have calculated the wealth inequality considering cumulative runs scored/session for each 18 IPL seasons and listed in Table \ref{Table1}. These $g$ and $k$ values are also plotted in Fig.\ref{fig2} as a line plot with 18 different points whose y-value is the wealth inequality and x-value is the corresponding IPL session. Here, the green line with triangular points represent the Gini index for the wealth inequality/1 season while the pink line represents the Kolkata index for the same.

%%%%%%%%%%%%%%%%%%% 3rd Figure %%%%%%%%%%%%%%%%%

\begin{figure}[ht]
    \centering % Center the figure
    % Do not include the .eps extension in the file name
    \includegraphics[width = 1.0 \textwidth]{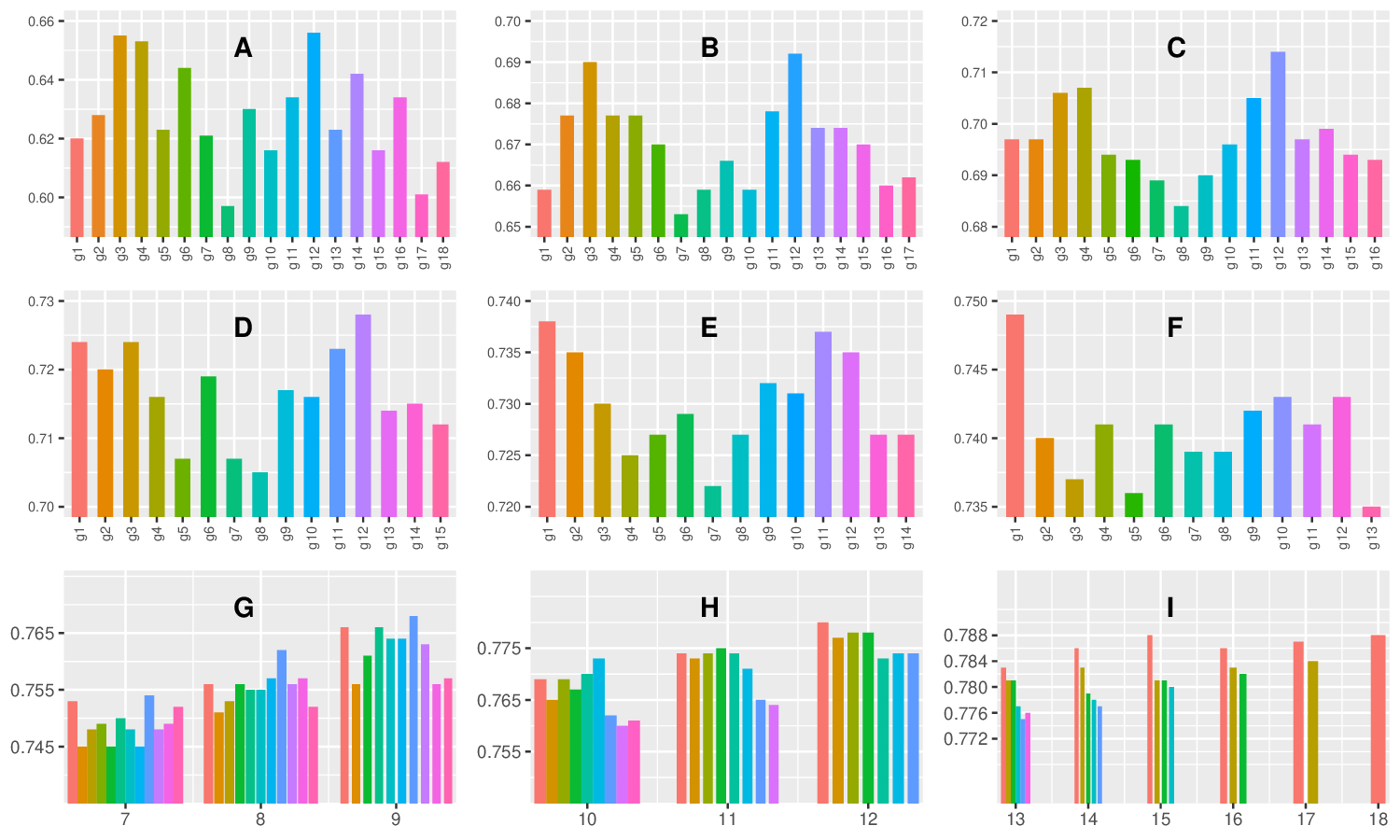} 
    \caption{Variation of the wealth inequality in terms of the Gini index, $g$ considering cumulative runs scored in 1 season, 2 seasons, .... up to whole period, i.e. 18 IPL seasons. At the same time each sub figure shows how wealth fluctuates for different choices of considering cumulative runs up to some particular successive seasons out of total 18 seasons. Here, y-axis reflects the value of $g$. Fig.\ref{fig3}A represents inequality considering cumulative runs scored in 1 IPL season. For 18 different seasons, $g$ has 18 different values represented by colour bar. Here x-axis represents number of choices the seasons were considered. Fig.\ref{fig3}B represents $g$ value considering cumulative runs for two consecutive seasons. Out of 18 seasons, one can choose 2 consecutive seasons 17 different ways, there are 17 different colour bar. Fig.\ref{fig3}C represents $g$ value for cumulative runs up to 3 consecutive seasons. Out of 18 seasons this can be chosen in 16 different ways. Fig.\ref{fig3}D represents $g$ for cumulative runs up to 4 consecutive sessions and here 15 different combinations of such seasons are possible. Each colour bar represents the value of the Gini index for each choice. Fig.\ref{fig3}E represents Gini index considering cumulative runs for 5 consecutive seasons which can be chosen as 14 different ways and each bar represents corresponding $g$ values. Fig.\ref{fig3}F is for wealth inequality considering cumulative runs up to 6 consecutive seasons. There are 13 such seasons combinations. We have shown $g$ values for 7 to 9 consecutive sessions in Fig.\ref{fig3}G. Here 12 colour bar above 7 represents different $g$ values for 12 different combination of 7 consecutive seasons. The same is above 8 and 9. In the same way Fig.\ref{fig3}H represents different values of $g$ for each different choice of cumulative runs up to 10, 11 and 12 consecutive IPL seasons and Fig.\ref{fig3}I is the same for cumulative runs up to 13, 14, 15,....,18 consecutive sessions of IPL. In short, we have shown $g$ values for wealth inequality considering cumulative runs for 1 season to 6 consecutive seasons in Fig.\ref{fig3}A to Fig.\ref{fig3}F. While Fig.\ref{fig3}G represents the same for 7 consecutive sessions to 9 consecutive sessions in one plot and Fig.\ref{fig3}I represents the same for 13 consecutive to 18 consecutive seasons. This has been done so that one can understand how $g$ values varies with different choices of some specific consecutive sessions. Studying the numbers along y-axis we can understand how $g$ fluctuates for each choices of successive IPL seasons. Here, we can note that as we have considered more and more IPL seasons to calculate cumulative runs, variation in $g$ becomes less with each different choice of IPL sessions. Again wealth inequality gradually increases when we calculate the cumulative runs considering more and more successive sessions.}
    \label{fig3}
\end{figure}

%%%%%%%%%%%%%%  End of 3-rd figure %%%%%%%%%%%

\subsection{Change of wealth inequality over time}

Wealth is not limited up to a certain period of time. It grows over time. Cumulative runs which indicates wealth of a batter also increases as batters plays more and more seasons. So, if we consider cumulative runs up to more and more IPL seasons, we can also study how inequality changes with growing wealth. At first we choose two consecutive sessions. Out of total 18 seasons, one can choose two consecutive seasons in 17 different ways and there are 17 pairs of $g$ and $k$. One can find how $g$ changes with each different choice of consecutive sessions in Fig.\ref{fig3}. In Fig.\ref{fig3}A, we have shown how $g$ varies for 18 different choices if we consider cumulative runs up to 1 season only as discussed earlier. Fig.\ref{fig3}B represents the present case, i.e., how $g$ varies with 17 different choices of consecutive season if we consider cumulative runs up to 2 consecutive seasons. As we consider more and more successive seasons, number of different choice of considering seasons becomes lower as total number of IPL seasons played till now is 18. Fig.\ref{fig3}C to Fig.\ref{fig3}F represent the same if we consider cumulative runs up to 3, 4, 5 and 6 consecutive sessions. In Fig.\ref{fig3}G,  we have plotted variation of wealth inequality in terms of $g$ with cumulative runs up to 7, 8 and 9 consecutive seasons. Different colour bar above 7 represent value of $g$ for different choices of 7 consecutive seasons. The same rule has been followed for season 8 and 9. As with increasing seasons variation in $g$ with different choices decreases, so instead of plotting three different graphs as earlier cases, we have shown them in one graph. Following the same technique we have calculated and plotted wealth inequality in terms of the Gini index considering cumulative runs up to 10, 11 and 12 consecutive seasons in Fig.\ref{fig3}H. Here also different colour bars above 10 represent values of $g$ for different choice of considering 10 consecutive seasons. We followed the same for 11 and 12 consecutive sessions in the same plot. In the last plot of Fig.\ref{fig3}, i.e. in  Fig.\ref{fig3}I, We have plotted the wealth inequality considering the cumulative runs made by batters in 13, 14, 15, 16, 17 and 18 consecutive IPL seasons. The numbers along the x-axis denotes the successive seasons we have considered and the different colour bars above them denotes value of $g$ for different choices of consecutive seasons out of total 18 IPL seasons. These plots describes how $g$ value fluctuates for different choices under a specific successive IPL seasons considered. 

\begin{figure}[ht]
    \centering % Center the figure
    % Do not include the .eps extension in the file name
    \includegraphics[width = 1.0 \textwidth]{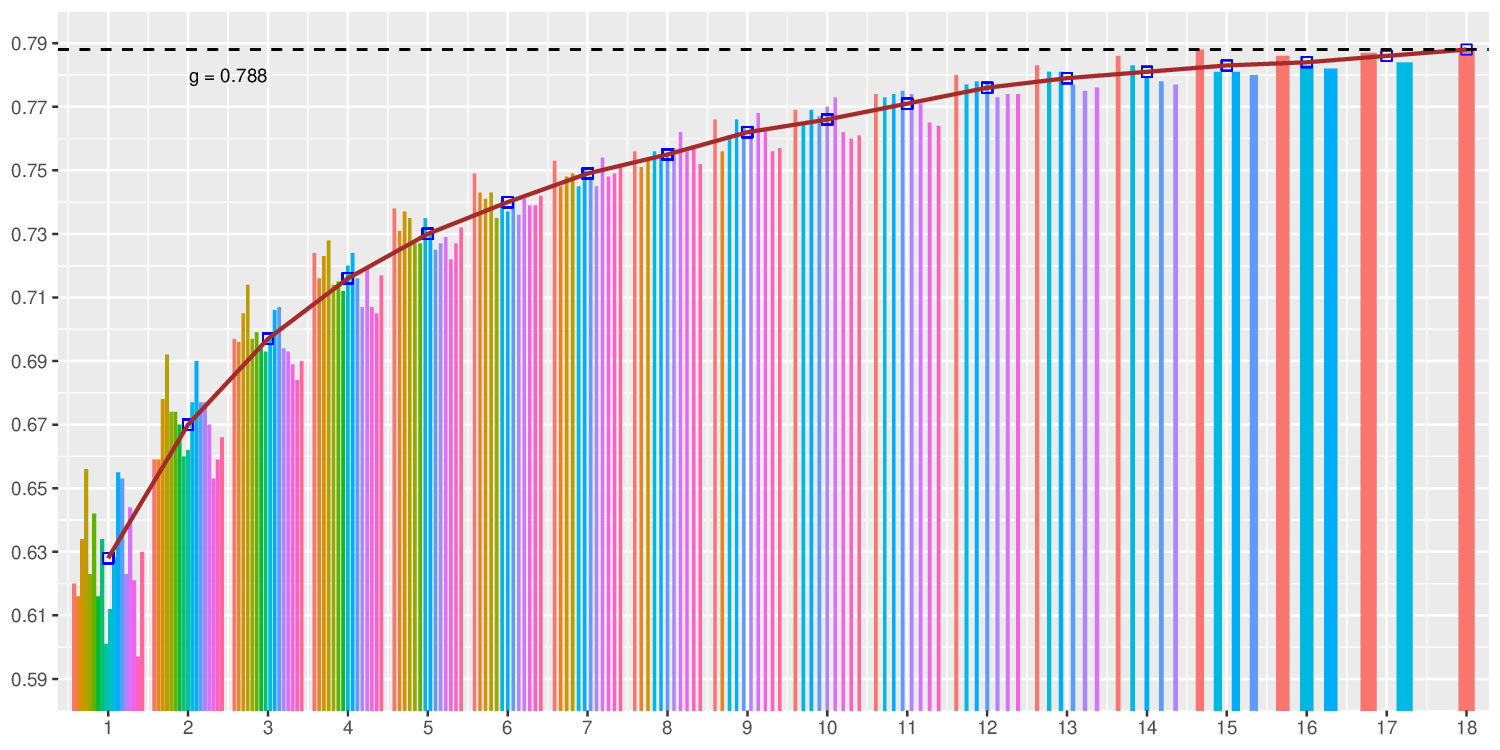} 
    \caption{Increase of wealth inequality in terms of the Gini index as we considered cumulative runs up to more and more successive seasons. y-axis shows the wealth inequality in Gini Index. Numbers along the x-axis represent successive seasons for which we considered cumulative runs scored by batters. Thus different colour bar plots above each session represent values of wealth inequality in terms of the Gini index for different choices we can make when considering respective successive seasons out of total 18 IPL seasons. Here we observe that, with the increase in number of successive sessions wealth inequality increases. When we consider all IPL seasons, the wealth inequality is around 0.8. The nature of the increase in inequality is asymptotic in nature. To show the asymptotic nature clearly, we have marked the average inequality for cumulative runs up to successive seasons by blue square points. The line plot joining the average $g$ values touches the asymptotic line $g = 0.788 $ at the end of 18 successive seasons.}
    \label{fig4}
    
\end{figure}

\begin{figure}[ht]
    \centering % Center the figure
    % Do not include the .eps extension in the file name
    \includegraphics[width = 1.0 \textwidth]{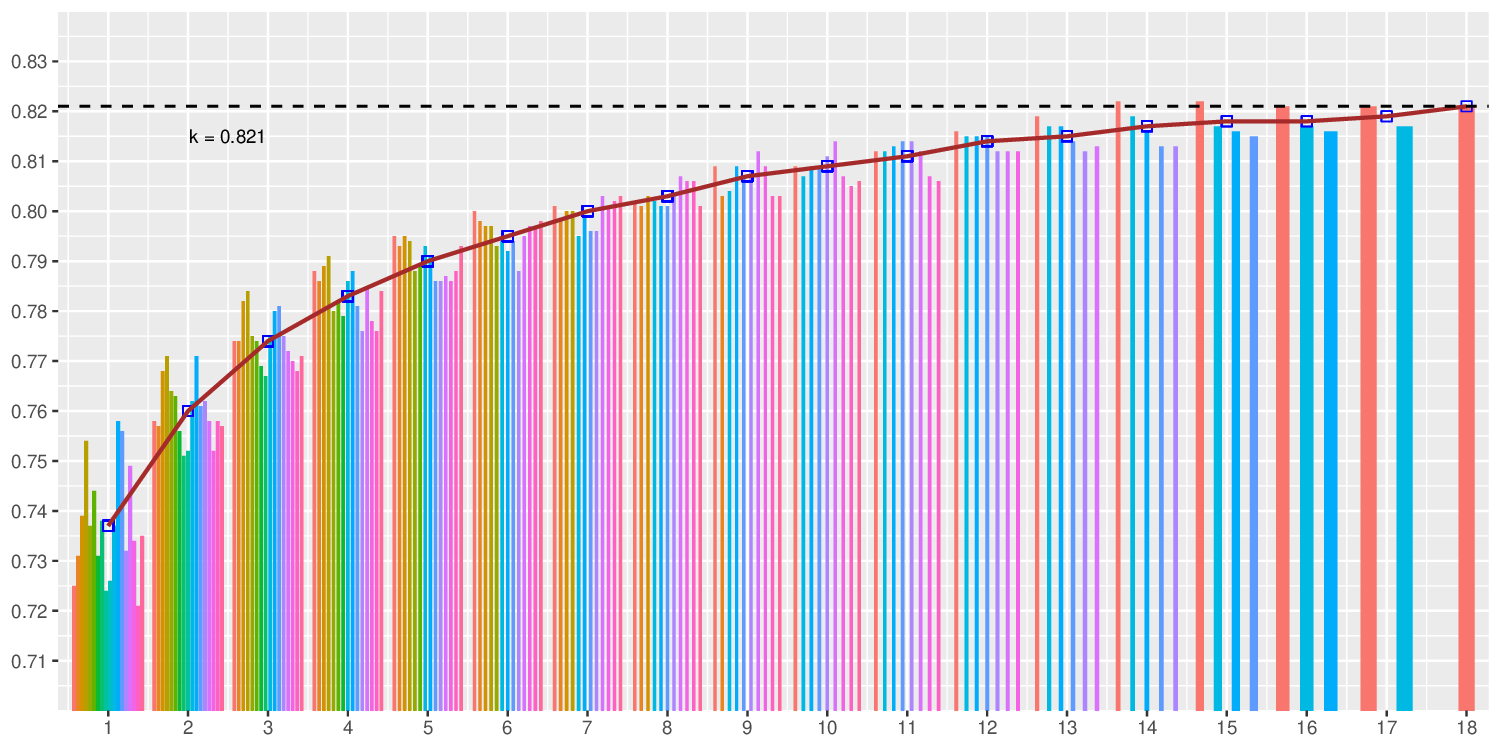} 
    \caption{Increase of wealth inequality in terms of the Kolkata index as we considered cumulative runs up-to more and more successive session. Here y-axis shows the wealth inequality in Kolkata Index. Numbers along the x-axis represent successive sessions for which we considered cumulative runs scored by batters. Thus different colour bar plots above each session represents values of wealth inequality in terms of the Kolkata index for different choices we can make when considering respective successive seasons out of total 18 IPL seasons. Here we observe that, with the increase in number of successive sessions wealth inequality increases. When we consider all IPL seasons, the wealth inequality is around 0.82 and it is the approximate asymptotic value of inequality in terms of the Kolkata index.}
    \label{fig5}
\end{figure}

To show how wealth inequality in terms of the Gini index changes as we consider more and more successive seasons, we have plotted the whole data of Fig.\ref{fig3} in Fig.\ref{fig4}. This plot clearly indicates that increase in wealth inequality is asymptotic in nature. At first, the increase in inequality is faster as we consider more and more seasons. However, the increase slows down and stabilizes around Gini index 0.8 at the end of 18 IPL seasons. This means, with increase in wealth over time, inequality also increases. So, we can infer that as sessions increases a few number of batters are able to continue in teams. As a consequence, after 18 IPL session, if we analyze the cumulative runs made by different batters, a few batters make lion's share of the total runs made in IPL history. As the Gini index is not linked directly with Pareto's law, we have also computed wealth inequality in terms of the Kolkata index also. Fig.\ref{fig5} represents the same as we have just discussed but in terms of the Kolkata index. Here, we observe that the asymptotic value of the Kolkata index is around 0.82. This ensures that over time nearly 20\% batters own 80\% of the total runs scored so far. This is exactly the same as depicted in the Pareto principle.

%%%%%%%%%%%%%%% End of section 4 %%%%%%%%%%%%%%

%%%%%%%%%%%%%%%% 4th Figure %%%%%%%%%%%%%%%%%%%%%%

%%%%%%%%%%%%%%%%% End of 4-th figure %%%%%%%%%%

%%%%%%%%%%%%%%%%% 5th Figure %%%%%%%%%%%%%%%%%

%%%%%%%%%%%%%%% End of 5th Figure %%%%%%%%%%%%%%

%%%%%%%%%% 5th section %%%%%%%%%%%%%%%%%%%%%%

\section {Presentation of calculated data of income and wealth inequality}

In the previous section we have described in detail how we have chosen required data to calculate income and wealth inequality with the help of runs scored by batters in IPL. Based on this, we have plotted 4 different plots. Fig.\ref{fig2} shows the variation of the income inequality in terms of the Gini index, $g$ and the Kolkata index, $k$ with different IPL seasons. Here, each number along the x-axis indicates different IPL season. The y-axis describes the value of the corresponding inequality. It also describes how wealth inequality based on cumulative runs scored by batters considering 1 IPL season, varies with each session in terms of $g$ and $k$. 

Fig.\ref{fig3} is a combination of 9 plots labeled with different English letters. Different plots show how $g$ values related to wealth inequality increase  if cumulative runs are counted for more and more successive IPL seasons. Wealth inequality reaches maximum when we count cumulative runs for the entire period of IPL, i.e. all 18 IPL seasons till date. At the same time, different plots of Fig.\ref{fig3} describe how $g$ values of the wealth inequality vary with different choice of successive IPL seasons out of total 18 seasons for some specific successive seasons under study. Suppose, we are considering wealth inequality in cumulative runs scored up to 2 successive IPL seasons. Now, out of total 18 seasons, we can choose 2 consecutive seasons 17 different ways. For each choice, we can calculate wealth inequality. So, in this case we have 17 different values of wealth inequality which are not same. Here we have studied how these fluctuations among inequality of different choices varies with more and more successive sessions considered. As, number of choices are more when successive seasons under study are few, we have plotted fluctuations of $g$ for each different choices when successive seasons under study are 2 to 6 in different plots. Plot Fig.\ref{fig3}A shows the fluctuations in $g$ describing wealth inequality, when cumulative runs were counted up to two successive seasons. As we have said we can choose 2 such successive season out of 18 total season in 17 different ways, there are 17 different colour bars in Fig.\ref{fig3}A each representing value of $g$ for each different choices. Along the x-axis we have marked each bar by different notations to denote corresponding $g$ values with different choices. The numbers along the y-axis describes the magnitude of inequality. Fig.\ref{fig3}B, Fig.\ref{fig3}C, Fig.\ref{fig3}D, Fig.\ref{fig3}E and Fig.\ref{fig3}F represents the same for 3, 4, 5, 6 and 7 successive seasons. Number of choosing successive seasons out of total 18 seasons reduces as we consider more and more successive seasons. So, instead of plotting fluctuations in $g$ individually for each considered number of successive seasons as in previous cases, we have plotted the same for 3 cases in Fig.\ref{fig3}G. Here the numbers along the x-axis represent how many successive seasons we have considered and the numbers of colour bars denotes how many choices we can make to choose these successive seasons out of 18 seasons. The height of each bar represents corresponding values of $g$. The same has been plotted in Fig.\ref{fig3}H for 10, 11 and 12 successive seasons. In Fig.\ref{fig3}I, we have plotted the same for 13 to 18 successive seasons.

Although all the plots of Fig.\ref{fig3} represent how wealth inequality increases with more and more inclusion of successive seasons of IPL, it does not explore how the increase occurs, i.e., whether this increase is random or gradual. To understand this feature, we have also plotted all the subplots of Fig.\ref{fig3} in one plot Fig.\ref{fig4}. It is a group bar plot where the x-axis represents the number of sessions we have considered to calculate cumulative runs. Observing Fig.\ref{fig4} we find that there are many colour bars above each number. The height of these individual bars represents the value of the wealth inequality in terms of $g$ for each choice of selecting successive seasons. As the number of seasons under consideration increases, the number of different choices decreases. So, each group of bars is related to the number of seasons that we have considered for computing cumulative runs. Here we can observe that the overall height of the grouped bars gradually increases with the more and more IPL seasons considered to calculate the cumulative runs. Each blue square point above each group of bars represents the corresponding average value of $g$. The line plot which connects these square points shows an asymptotic nature with final value of wealth inequality around 0.79 in terms of the Gini index. 

We have already stated that the Gini index does not directly depict the Pareto principle, rather, the Kolkata index is directly related to it. So, to get an idea of the wealth inequality which we have estimated in view of cumulative runs scored by batters in the IPL, how much it resembles the Pareto principle, we have also plotted the same data, as stated above, in terms of the Kolkata index in Fig.\ref{fig5}. It is similar to Fig.\ref{fig4} but wealth inequality has been expressed in terms of the Kolkata index. Here, also the blue squares above each group of bars represent the corresponding average value. The asymptotic line plot that connects all these average values approaches the limiting value around $k = 0.82$. So, we can definitely comment that here also 82\% of the cumulative runs made in all IPL seasons were scored by 18\% of the batters played in the entire IPL seasons. This is almost in accordance with the Pareto principle.

 %%%%%%%%%% End of section V %%%%%%%%%%%%%%%%%%%%

\section{Discussions and Conclusions}
In the previous section, we have presented the calculated data related to this study in terms of 4 different plots. Of them, only Fig.\ref{fig2} is related to the income and wealth inequality, and others are related only to the wealth inequality. In this section, we critically interpret the above plots and compare the results with known information on income and wealth inequality in reality. 
\subsection{Income Inequality}
Careful observation of Fig.\ref{fig2} reveals the following:

The Gini indices that represent income inequality based on runs scored in an innings, i.e., $g_{inc}$, fluctuate between 0.54 and 0.57. The same is true for the Kolkata indices, which vary between 0.69 and 0.71. This indicates that throughout the IPL seasons, the income inequality remains more or less the same. It fluctuates very little around the average value of 0.557 in terms of the Gini index and 0.704 in terms of the Kolkata index. This means, 70\% of total income is shared by 30\% of people. Now, let us check income inequality around the world. Many trusted organizations publish data on country-wise income inequality each year \cite{7, 7a, 7b, 7c,7d, 7f}. Here, we observe that income inequality varies from 0.24 to 0.64 in terms of the Gini index. We have already discussed that income inequality depends upon various policies taken by the respective countries to mitigate the differences between "the have and have-nots". Policies to reduce income inequality focus on redistribution through progressive taxation, strengthening labor rights (minimum wage), expanding access to education and training, and direct income transfers (social safety nets). Key strategies include taxing high earners and wealth, improving public service access, and supporting marginalized groups to improve equitable opportunity. India also follows these social welfare policies and has an income inequality of around 0.26 (Gini index). However, South Africa and most African countries have high income inequality (around 0.6 in Gini). 

In our analysis, average value of the Gini index is around 0.56 which is near the upper limit of income inequality observed in the real world. In addition, there is no mitigation measure that will reduce inequality. Instead, organizers motivate players to perform their best, increasing inequality. So, overall income inequality in every session is nearly equal and close to the upper limit of inequality observed in the real world. 

The above observation is really amazing. Players from all over the country participate in the IPL tournament. In the whole IPL history, many old players have retired and many new players have joined. Moreover, the venue of the matches changed from time to time, but the overall income inequality estimated using runs scored by batters in the IPL, is approximately the same in every season. In addition, the income inequality we observed in each season of IPL is consistent with the upper limit of income inequality as seen in reality. Above all, in this analysis process, there is no ambiguity in the accuracy of the data. In the process, one can consider this to be the possible upper limit of income inequality, i.e., the Gini index around 0.6.

\subsection{Wealth Inequality}

Now, let us focus on wealth inequality. At first, we have calculated the wealth inequality considering cumulative runs for 1 session only. Starting from 2008, up to 2025, total of 18 IPL seasons have been played. Taking into account each session once, one can calculate 18 values of the Gini and Kolkata index which represent each session. These values have been listed in Table \ref{Table1} and plotted the same in Fig.\ref{fig2}. A careful observation of Fig.\ref{fig2} reveals that the Gini index for the wealth inequality/IPL season varies from 0.6 to  0.66. In this case, the average value of wealth inequality is around $g = 0.63$. The Kolkata index fluctuates between 0.72 and 0.76 with average value around 0.74, and these results are for cumulative run count up to one season. If we compare this wealth inequality with the income inequality, we observe that the wealth inequality is higher than the income inequality. Again, the fluctuation in wealth inequality around the average value is comparatively higher than the income inequality.

We have discussed earlier that wealth accumulates over time. Keeping this point in mind, if we consider cumulative runs up to more and more successive sessions of the IPL, the cumulative runs scored by the batters will increase. Now, let us check how inequality in wealth varies in the above process. It has been shown in Fig.\ref{fig3}, Fig.\ref{fig4} and Fig.\ref{fig5}.

Fig.\ref{fig3}A is nothing more than the bar graph presentation of the line graph given in Fig.\ref{fig2}. If we carefully observe Fig.\ref{fig3}B, we notice that $g$ varies from 0.65 to 0.69 for 17 different choices of 2 successive seasons out of 18 total seasons. The average value is approximately $g = 0.67$ as shown in Fig.\ref{fig4}. Comparing this with the wealth inequality calculated cumulative runs for 1 season, one can notice that the average value inequality has been increased but the span of fluctuation in different observations has been decreased. The same trend continues if we consider 3 successive seasons to calculate the cumulative runs. Fig.\ref{fig3}C informs that the fluctuation in $g$ in this case is from 0.685 to 0.715 and the average value is approximately $g = 0.70$ as in Fig.\ref{fig4}. As we go for more and more successive seasons, the fluctuation of the Gini index among different choices becomes smaller and smaller, and at the same time the average value increases. This has been shown in detail in all subplots of Fig.\ref{fig3}. However, the same nature of wealth inequality has been shown in Fig.\ref{fig4} which may be considered as a summary of Fig.\ref{fig3}. Here, we observe that with an increase in successive seasons, the average value of $g$ increases asymptotically and approaches $g = 0.79$. At the same time, fluctuations in $g$ between different choices become lower and lower. The same trend of wealth inequality in terms of the Kolkata index is shown in Fig.\ref{fig5}. The asymptotic value of wealth inequality in terms of the Kolkata index is around 0.82. So, we can say that 82\% of the total runs scored in the IPL have been scored only by 18\% of the total batters who participated in this tournament from the beginning. This is in par with the Pareto principle.

The above analysis reveals that as we consider more and more seasons to count the cumulative runs, although it increases cumulative runs made by different players but at the same time inequality also increases. This means that over time, as wealth increases among the families, but at the same time wealth inequality also increases, i.e., over time wealth accumulates more and more to a few households relative to others. Let us compare these outcomes with the real world data. Data of country-wise wealth inequality, in terms of the top 10\% population who hold how much share of total wealth is provided by some organizations \cite{8a, 8c, 9a}. In terms of the Gini index, global wealth inequality is around 0.80 \cite{8b, 14a, 14b}. In the process, we can conclude that the result we have obtained on wealth inequality based on the cumulative runs scored in IPL, shows more or less a similar result with the real world wealth inequality data.

\subsection{Conclusions}

In this study, we used runs scored by batters in the Indian Premier League (IPL) as an alternative way to calculate income and wealth inequality within society. In the absence of accurate real data, such types of proxy data prove to be useful for analyzing income and wealth inequality. Taking into account individual runs made by the batters in an innings as income and cumulative runs in seasons as wealth, this study shows that the calculated income and wealth inequality mirror the observed income and wealth inequality observed in real-world socio-economic system. In particular, the average income inequality calculated from the runs made in an innings in an IPL season is relatively the same for the entire IPL History, and this value of the income inequality aligns with the upper range of global income inequality. 

Furthermore, we have shown how wealth inequality increases over time as cumulative performance grows with the inclusion of more and more IPL seasons under consideration. Wealth inequality evolves asymptotically in the absence of any mitigating policies and approaches a limiting value which is consistent with empirical observations and the Pareto principle. The asymptotic values obtained ($g \approx 0.79$ and $k \approx 0.82$) indicate that the inequality tends to converge towards a universal upper bound under unregulated competitive conditions. These foundings highlight the potential of proxy based frameworks to mirror realistic socio-economic issues where accurate data are hard to collect. Again, this study explores all the essential features of inequality dynamics and provides a novel perspective for understanding the intrinsic limits of socio-economic inequality. 
%%%%%%%%%%%%%%%%%%%%%%%%%%%%%%%%%%%%%%%%%%%

%%%%%%%%%%%%%%%%%%%%%%%%%%%%%%%%%%%%

\appendix 

\section{Income and wealth inequality data}

Here we have listed the income inequality data in terms of the Gini and the Kolkata index measured from IPL data for each season. At the same time, it also shows the wealth inequality data calculated considering the cumulative runs scored by batters in one particular season.
\setlength{\tabcolsep}{14pt}
\renewcommand{\arraystretch}{1.2}
\begin{table}[ht]
    \centering
    
    \vspace{0.2 cm}
    \begin{tabular}{*7c}
    \hline
        Season &  \multicolumn{2}{c}{Income Inequality}  & \multicolumn{2}{c}{Wealth Inequality} \\ 
        ~ & $g_{inc}$ & $k_{inc}$ & $g_{wel}$ & $k_{wel}$ \\ \hline
        1 & 0.567 & 0.708 & 0.62 & 0.725 \\ \hline
        2 & 0.569 & 0.712 & 0.628 & 0.737 \\ \hline
        3 & 0.554 & 0.705 & 0.655 & 0.758 \\ \hline
        4 & 0.551 & 0.703 & 0.653 & 0.756 \\ \hline
        5 & 0.549 & 0.702 & 0.623 & 0.732 \\ \hline
        6 & 0.565 & 0.710 & 0.644 & 0.749 \\ \hline
        7 & 0.543 & 0.696 & 0.621 & 0.734 \\ \hline
        8 & 0.545 & 0.696 & 0.597 & 0.721 \\ \hline
        9 & 0.566 & 0.709 & 0.63 & 0.735 \\ \hline
        10 & 0.564 & 0.708 & 0.616 & 0.731 \\ \hline
        11 & 0.554 & 0.703 & 0.634 & 0.739 \\ \hline
        12 & 0.565 & 0.708 & 0.656 & 0.754 \\ \hline
        13 & 0.548 & 0.701 & 0.623 & 0.737 \\ \hline
        14 & 0.562 & 0.707 & 0.642 & 0.744 \\ \hline
        15 & 0.562 & 0.706 & 0.616 & 0.731 \\ \hline
        16 & 0.570 & 0.709 & 0.634 & 0.738 \\ \hline
        17 & 0.539 & 0.696 & 0.601 & 0.724 \\ \hline
        18 & 0.544 & 0.698 & 0.612 & 0.726 \\ \hline
    \end{tabular}
    \caption{Income and wealth inequality in terms of the Gini and the Kolkata index for each IPL seasons. Starting from the first season played in 2008, we have considered last IPL season, i.e., 18-th season played in 2025. Here $g_{inc}$ indicates the Gini value of income inequality measured using IPL data. $k_{inc}$ indicates the Kolkata index value of the same. $g_{wel}$ and $k_{wel}$ represents the calculated value of wealth inequality based on IPl data for a particular season in terms of the Gini and the Kolkata index.} 
    \label{Table1}
\end{table}

\end{document}